\documentclass[12pt]{article}
\usepackage{amsmath}
\usepackage{amssymb}
\usepackage{epsfig}
\usepackage{euscript}
\usepackage{fancybox}
\usepackage{color}

\def\mathswitchr#1{\relax\ifmmode{\mathrm{#1}}\else$\mathrm{#1}$\fi}

\newcommand {\pslash}{\hbox{$\not\hbox{\kern-2.3pt $p$}$}}

\usepackage{cite}

\usepackage{epic}

\def\alf1{ {\alpha\over\pi} }
\def\rQCED{{\rm QCED}}

\begin{document}
\begin{titlepage}
\begin{flushright}
{\bf CERN-PH-TH/2011-077}\\
{\bf BU-HEPP-11-02}\\
{\bf Apr., 2011}\\
\end{flushright}
\vspace{0.05cm}
 
\begin{center}
{\Large QCD for the LHC$^{\dagger}$}
\end{center}

\vspace{2mm}
\begin{center}
{\bf   B.F.L. Ward}\\
\vspace{2mm}
{\em Department of Physics,}\\
{\em Baylor University, Waco, Texas, USA}\\
{\em and}\\
{\em TH Unit, CERN, Geneva, Switzerland}\\
\end{center}

\vspace{5mm}
\begin{center}
{\bf   Abstract}
\end{center}
We discuss the new era of precision QCD as it relates to the physics requirements of the LHC for both the 
signal and background type processes. Some attention is paid to the issue of the theoretical 
error associated with any given theoretical prediction. In the cases considered, we present where the 
theory precision is at this writing and where it needs to go in order that it not impede the discovery 
potential of the LHC physics program. To complete the discussion, we also discuss possible paradigms 
the latter program may help us understand and some new developments that may play a role in achieving that respective understanding.
\\
\vskip 3mm
\centerline{Invited talk presented at the 2011 Epiphany Conference, Krakow, Poland}
\vskip 16mm
\vspace{10mm}
\renewcommand{\baselinestretch}{0.1}
\footnoterule
\noindent
{\footnotesize
\begin{itemize}
\item[${\dagger}$]
Work partly supported by US DOE grant DE-FG02-09ER41600. 
\end{itemize}
}

\end{titlepage}
 
\baselineskip=11pt 
\def\Kmax{K_{\rm max}}\def\ieps{{i\epsilon}}\def\rQCD{{\rm QCD}}
\renewcommand{\theequation}{\arabic{equation}}
\font\fortssbx=cmssbx10 scaled \magstep2
\renewcommand\thepage{}
\parskip.1truein\parindent=20pt\pagenumbering{arabic}\par

\section{\bf Introduction}\label{intro}\par
As the start-up of the LHC has precipitated the era of precision QCD, 
by which we mean
predictions for QCD processes at the total precision tag of $1\%$ or better,
it is appropriate for any discussion of the requirements on QCD for the LHC 
to set its framework by recalling, at least in generic terms, why we need 
the LHC in the first place. In the following discussion of the QCD for the 
LHC, we shall begin with such recollection. In this way, the entire effect 
of the effort required to realize precision QCD for the LHC in a practical 
way can be more properly assessed.\par
Thus, we ask, ``Why do we need the LHC?'' Many answers can be found in the 
original justifications for the colliding beam device and its detectors in Refs.~\cite{lhc,atlas,cms,lhcb,alice}. We will call attention to a particular
snap shot
of the latter discussions with some eye toward the requirements of precision QCD from the theoretical standpoint. More precisely, the LHC is a crucial step toward resolving fundamental outstanding issues in elementary particle physics:
the big and little hierarchy problems,
the number for families,
the origin of Lagrangian fermion (and gauge boson) masses,
baryon stability,
the union of quantum mechanics and general theory of relativity,
the origin of CP violation,
the origin of the cosmological constant $\Lambda$,
dark matter,~
$\cdots$ .
Much theory effort has been invested in the ``New Physics'' (NP) 
that would seem to be needed to solve all of these outstanding issues, 
that is to say, in the physics beyond the Standard Model 't Hooft-Veltman 
renormalized
Glashow-Salam-Weinberg EW $\times$ Gross-Wilczek-Politzer QCD theory that
seems to describe the quantum loop corrections in the measurements of 
electroweak and strong interactions at the shortest distances 
so far achieved in laboratory-based experiments.\par
We mention that superstring theory~\cite{gsw,pol} solves everything 
in principle but has trouble in practice: for example it has more 
than $10^{500}$
candidate solutions for the vacuum state~\cite{stringvac}. The ideas in 
superstring theory have helped to motivate many so-called string inspired 
models of NP such as~\cite{bsm} string-inspired GUTs, large extra dimensions, Kaluza-Klein excitations, ... . 
We list supersymmetric extensions of the SM, such as the 
MSSM and the CMSSM~\cite{bsm}, as separate proposals from 
superstring motivated ideas, as historically this was the case. Modern 
approaches to the dynamical EW symmetry  breaking 
(technicolor) such as little Brout-Englert-Higgs models~\cite{bsm}, 
obtain as well. The list is quite long and LHC will help us shorten it, no 
doubt.\par
Perhaps, one of the most provocative ideas is the one which some superstring 
theorists~\cite{stringvac} invoke to solve the problem of the large number of 
candidate superstring vacua: the anthropic principle, by which the solution 
is the one that allows us to be in the state in which we find ourselves. 
In the view of some~\cite{susskd}, this would be the end 
of reductionist physics as we 
now know it. Can LHC even settle this discussion? Perhaps.\par
More recently, even newer paradigms are emerging. In Ref.~\cite{giud-dvli},
the UV limit of theories such as quantum gravity is solved by the dynamical generation of non-perturbative large distance excitations called classicalons, which provide the necessary damping of the naively divergent UV behavior. When discussed in general terms, possible new signatures for the LHC obtain~\cite{giud-dvli}.\par
In Ref.~\cite{leguts}, the $E_8\times E_8\equiv E_{8a}\times E_{8b}$ symmetry group suggested by the heterotic string theory~\cite{hstrg} is abstracted to apply to the fundamental symmetry group physics for GUT's and it is shown that, if one presumes that the known light leptons are in the 3-families of SO(10) \underline{16}'s with three sets of new quarks, $\{u',d';c',s';t',b'\}$ while the known quarks are in three families of
SO(10) \underline{16}'s with three new sets of heavy leptons $\{\nu_{\ell'},\ell', \; \ell'= e',\;,\mu',\;\tau'\}$, where the two sets of three SO(10) families can either be generated by breaking the  $E_{8a}\times E_{8b}$ such that the first(second) set transforms non-trivially only under $E_{8a}$ ($E_{8b}$) or
be generated by having both sets of three families transform non-trivially under $E_{8a}$, leaving open the possibility of an unspecified number of families, as yet unseen, to transform under non-trivially under $E_{8b}$. The proton is stable for purely kinematic reasons -- all the leptons to which it could decay are too heavy for the decay to occur. The mixing matrix for the low energy EW gauge bosons from the GUT scale breaking of
the  $E_{8a}\times E_{8b}$ symmetry down to the Standard Model gauge group then allows the GUT scale $M_{GUT}$ to obtain at $\lesssim 200$TeV, in reach of
the VLHC colliding beam device as discussed in Refs.~\cite{vlhc}. Many of the new heavy quarks and leptons in this paradigm could already be visible at the LHC.\par
If one does not use string theory for the unification of the EW and QCD theories with quantum gravity, then  one needs a remedy for the UV sector of quantum gravity. Recently, in addition to the ideas in Ref.~\cite{giud-dvli}, more progress has been made on solving this problem in the context of local Lagrangian field theory methods~\cite{reuter-picac-litm,rqg,kreimer}. Specifically, following the suggestion by Weinberg~\cite{wein1} that quantum gravity might have a non-trivial UV fixed point, with a finite dimensional critical surface
in the UV limit, so that it would be asymptotically safe with an S-matrix
that depends on only a finite number of observable parameters, 
in 
Refs.~\cite{reuter-picac-litm} 
strong evidence has been calculated
using Wilsonian~\cite{kgw} field-space exact renormalization group methods to support
Weinberg's asymptotic safety hypothesis for the Einstein-Hilbert theory.
In a parallel but independent development~\cite{rqg}, we have shown~\cite{bw2i} that the extension of the amplitude-based, exact resummation theory of Ref.~\cite{yfs} to the Einstein-Hilbert theory leads to UV-fixed-point behavior for the dimensionless gravitational and cosmological constants with the bonus that the resummed theory is actually UV finite when expanded in the resummed propagators and vertices's to any finite order in the respective improved loop expansion.
We refer to the resummed theory as resummed quantum gravity. 
In addition, more evidence for Weinberg's asymptotic safety behavior has been calculated using causal dynamical triangulated lattice methods in Ref.~\cite{ambj}\footnote{We also note that the model in Ref.~\cite{horva} realizes many aspects
of the effective field theory implied by the anomalous dimension of 2 at the
UV-fixed point but it does so at the expense of violating Lorentz invariance.}.
At this point, there is no known inconsistency between our analysis
and those of the Refs.~\cite{reuter-picac-litm,ambj} or
the leg renormalizability arguments in Ref.~\cite{kreimer}.
We note further that, in Refs.~\cite{reuter1,reuter2}, it has been argued that the approach in Refs.~\cite{reuter-picac-litm} 
to quantum gravity
may indeed provide a realization\footnote{The attendant choice of the scale $k\sim 1/t$ used in Refs.~\cite{reuter1,reuter2} was also proposed in Ref.~\cite{sola1}.} of the successful
inflationary model~\cite{guth,linde} of cosmology
without the need of the as yet unseen inflaton scalar field: the attendant UV fixed point solution
allows one to develop Planck scale cosmology that joins smoothly onto
the standard Friedmann-Walker-Robertson classical descriptions so
that then one arrives at a quantum mechanical 
solution to the horizon, flatness, entropy
and scale free spectrum problems. In Ref.~\cite{bw2i}, we have shown
that, in the new
resummed theory~\cite{rqg} of quantum gravity, 
we recover the properties as used in Refs.~\cite{reuter1,reuter2} 
for the UV fixed point of quantum gravity with the
added results that we get ``first principles''
predictions for the fixed point values of
the respective dimensionless gravitational and cosmological constants
in their analysis. In Ref.~\cite{bw-lambda}
we carry the analysis one step further and arrive at a prediction for 
the observed cosmological constant 
$\Lambda$, $\rho_\Lambda\cong (2.400\times 10^{-3}eV)^4$, 
in the
context of the Planck scale cosmology of Refs.~\cite{reuter1,reuter2},
which is reasonably close to the observed 
value~\cite{cosm1,pdg2008} $(2.368\times 10^{-3}eV(1\pm 0.023))^4$.\par
It follows that the new paradigms, which we have illustrated 
admittedly only in part in a limited way to set the stage of our 
discussion here, must be taken seriously in analyzing the new LHC data. 
In particular, we must be able to distinguish higher order SM processes 
from New Physics and we must be able to probe New Physics precisely to 
distinguish among different New Physics scenarios. This necessitates the 
era of precision QCD for the LHC.\par
Our discussion is organized as follows. We first discuss in the next Section
the issue of QCD at high energies from the standpoint of precision 
theory. Section 3 deals with applications of such theory to the LHC scenario 
and concludes with a look toward the future.\par
\section{QCD at High Energies}
At high energies when we have sufficiently large momentum transfer 
interactions, such as we have at the hard scattering processes at the LHC, 
we have the master formula for the respective fully differential cross 
sections as 
\begin{equation}
d\sigma =\sum_{i,j}\int dx_1dx_2F_i(x_1)F_j(x_2)d\hat\sigma_{\text{res}}(x_1x_2s)
\label{bscfrla}
\end{equation}
using a standard notation so that the $\{F_j\}$ and 
$d\hat\sigma_{\text{res}}$ are the respective parton densities and 
reduced hard differential cross section where we indicate the that latter 
has been resummed
for all large EW and QCD higher order corrections in a manner consistent
with achieving a total precision tag of 1\% or better for the total 
theoretical precision of (\ref{bscfrla}) as we discuss in more detail 
presently - this latter precision tag will be our definition of precision 
QCD theory. See Refs.~\cite{qced} where an example of such simultaneous 
QCD$\times$EW 
resummation is presented
- such resummation will be reviewed briefly in the following as well 
in the interest of completeness\footnote{For an alternative approach to
such simultaneous QCD$\times$EW resummation, see Refs.~\cite{ncrsni}}.\par
At high energies, we may have the hadron-hadron colliding beam paradigm, 
such as what we have at the LHC and at the Tevatron, the $e^+e^-$ colliding 
beam paradigm, as it is proposed now for the ILC, and the lepton-hadron 
colliding beam paradigm, as we had until recently at HERA. How do we assess 
the precision
of a theoretical result for (\ref{bscfrla}) in these paradigms? The
respective theoretical
precisions, $\Delta\sigma_{\text{th}}$, can be decomposed as follows:
\begin{equation}
\begin{split}
\text{Hadron-Hadron:}\; \Delta\sigma_{\text{th}}&= \Delta F \oplus\Delta\hat\sigma_{\text{res}}\nonumber\\
e^+e^-:\; \Delta\sigma_{\text{th}}&= 0\oplus\Delta\hat\sigma_{\text{res}}
\end{split}
\label{eqdecomp1}
\end{equation}
where the lepton-hadron case is covered by the hadron-hadron case if we 
interpret the parton density theory error, $\Delta F$, as that for just 
one factor of the $F_j$ in the cross section accordingly and where we 
note that, in the $e^+e^-$
high energy colliding beam case the analoga of the $\{F_j\}$ can be computed,
on an event-by-event basis by MC methods as a part of the resummed 
cross section~\cite{sjward}, $\hat\sigma_{\text{res}}$, so that we can 
set the analog of
$\Delta F$ to zero accordingly.\par
We stress that the theoretical precision, $\Delta\sigma_{\text{th}}$, 
validates the application of a given theoretical prediction to precision 
experimental observations, for the discussion of backgrounds for 
both SM and NP studies and
for the signals for both SM and NP studies, and more specifically
for the overall normalization
of the cross sections in such studies. NP can be missed if a calculation
with an unknown value of $\Delta\sigma_{\text{th}}$ 
is used to assess theoretical expectations 
for such studies. This point cannot be emphasized too much.\par
Here, we define $\Delta\sigma_{\text{th}}$ as the 
total theoretical uncertainty coming from the physical precision 
contribution and the technical precision contribution~\cite{jadach-prec}:
the physical precision contribution, $\Delta\sigma^{\text{phys}}_{\text{th}}$,
arises from such sources as missing graphs, approximations to graphs, 
truncations,....; the technical precision contribution, 
$\Delta\sigma^{\text{tech}}_{\text{th}}$, arises from such sources as 
bugs in codes, numerical rounding errors,
convergence issues, etc. The total theoretical error is then 
generically given by
\begin{equation}
\Delta\sigma_{\text{th}}=\Delta\sigma^{\text{phys}}_{\text{th}}\oplus \Delta\sigma^{\text{tech}}_{\text{th}}.
\end{equation}
The desired value for $\Delta\sigma_{\text{th}}$ depends on the  specific
requirements of the observations. As a general rule, one would 
like that $\Delta\sigma_{\text{th}}\le f\Delta\sigma_{\text{expt}}$, 
where $\Delta\sigma_{\text{expt}}$ is the respective experimental error
and $f\lesssim \frac{1}{2}$ so that
the theoretical uncertainty does not significantly affect the 
analysis of the data for physics studies in an adverse way.\par
For illustration we note the following examples that have obtained.
At the Tevatron one generally had the luminosity experimental uncertainty~\cite{fnal-lum} of $\Delta\sigma^{\text{norm}}_{\text{expt}}\cong 6-7\%$
so that theoretical predictions for cross sections at the level
of $\Delta\sigma_{\text{th}}\sim 10\%$ were acceptable in general. At LEP1,
the observation of 20M $Z$'s necessitated that the normalization error
from the theoretical cross section~\cite{sjward} was 
$\Delta\sigma^{\text{norm}}_{\text{th}}=0.061\%$ or better.
What do we need for the LHC physics in this context?\par

\section{Applications of Precision QCD for LHC Physics}

When we consider the LHC, we already see the effect of much better detectors
and much larger statistics compared to the Tevatron for example. Indeed, 
already, the experiments~\cite{lhcc-lum} are reporting a normalization 
error from experiment at the 3-4\% level, with the expectation~\cite{lhc-lum} 
that 1-2\% will be achieved. This defines a new set of goals for the 
theoretical uncertainty in QCD calculations for the LHC.\par
Specifically, the goals for the theoretical uncertainty 
$\Delta\sigma_{\text{th}}$for precision QCD calculations 
(henceforward we take it as understood that
we include the corresponding higher order EW and mixed EW$\otimes$QCD 
corrections as well) can be illustrated as follows:
for the so-called standard candle processes, we have 
\begin{equation}
\begin{split}
\text{Single}\; Z, W \;\text{production}:&\;\; \Delta\sigma_{\text{th}}\lesssim 1\%\\
t\bar{t} \;\text{production}:&\;\;\Delta\sigma_{\text{th}}\lesssim 1\%.
\end{split}
\label{eqszw}
\end{equation}
For the Les Houches~\cite{leshouches} list:
\begin{equation}
2\rightarrow n \;\text{processes to}\; {\cal O}(\alpha_s),\; n\ge 3: \;\;\Delta\sigma_{\text{th}}\cong 10\%.
\label{eqlesh}
\end{equation}
Exactness of the theoretical results is essential to have any chance of achieving these goals in a practical way.
What is the current state-of-the-art(SOTA) for published results on such goals?
\par
\subsection{SOTA for $\Delta\sigma_{\text{th}}$ for LHC Physics}
There has been significant progress on the goals outlined for $\Delta\sigma_{\text{th}}$. We now summarize some of this progress\footnote{We apologize if we omit some of references that should just as well be cited but we had to make some choices due to the limitations of space for this report.}.
On single $Z$ and $W$ production, the situation has been analyzed in Ref.~\cite{syost1} where the values found for $\Delta\sigma_{\text{th}}$ are as follows:
for single $Z$ production at $14$TeV, using a standard 
notation for effects from Ref.~\cite{syost1}, $\Delta\sigma_{\text{th}}=(4.91\pm0.38)\%=(2.45\pm0.73)\%(QCD+EW)\oplus 4.11\%(PDF)\oplus (1.10\pm0.44)\%(QCD Scale)$ and
for single $W^+[W^-]$ production at the same cms energy the corresponding value is $\Delta\sigma_{\text{th}}=(5.05\pm0.58)\%[(5.24\pm0)\%]$. The error due to the PDF uncertainty shown here
quantifies the type of value we have for the error $\Delta F$
in (\ref{eqdecomp1}). We see that in all three cases, there is still considerable effort that remains to be done to reach the goals presented in (\ref{eqszw}).\par
\begin{figure}[h]
\begin{center}
\includegraphics[width=80mm]{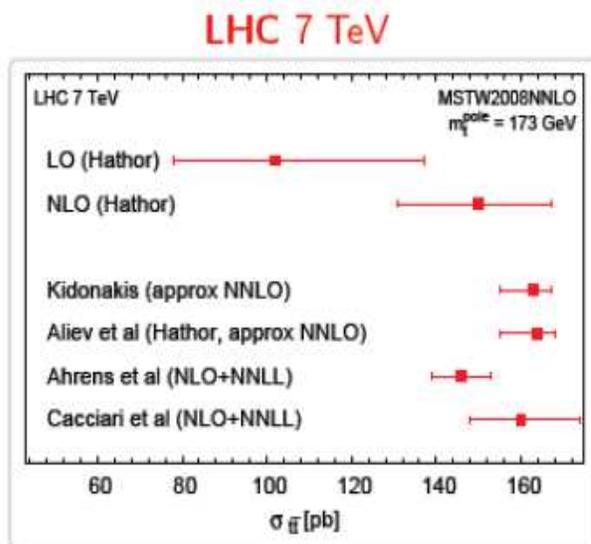}
\end{center}
\caption{\baselineskip=11pt  Results on $t\bar{t}$ production at the 
LHC as reviewed in Ref.~\cite{gslm}.}
\label{fig-gslm1}
\end{figure}

For $t\bar{t}$ production, the situation was recently reviewed by Salam in 
Ref.~\cite{gslm} as we reproduce here in Fig.~\ref{fig-gslm1}. The results 
shown do not contain any contribution from the respective PDF uncertainty, what we referred to as above as $\Delta F$, so that the total value of $\Delta_{\text{th}}$ is
$\Delta_F\oplus\Delta\sigma^{\text{rest}}_{\text{th}}$, where here $\Delta\sigma^{\text{rest}}_{\text{th}}$ refers to the errors shown in Fig.~\ref{fig-gslm1}
for cited calculation. If we use the basic estimator of the actual error via
~\cite{pdg2008} the standard formula ($\bar\sigma$ is the usual naive average)
\begin{equation}
\Delta_{\text{th}}^2\cong \frac{1}{N-1}\sum_{i=1}^{N}(\sigma_i-\bar\sigma)^2
\end{equation}
for the results in Fig.~\ref{fig-gslm1}, we arrive at the optimistic result
$\Delta_{\text{th}}\cong 4.8\%$ for the current SOTA for $t\bar{t}$ production
at the LHC at $7$TeV. This again is significantly larger than the goal in (\ref{eqszw}).\par
For the $2\rightarrow n,\; n\ge 3$ processes, the applications are 
to backgrounds to NP
and to more precision tests for the SM processes, wherein the ${\cal O}(\alpha_s)$ correction is essential. There has been great progress in achieving these
${\cal O}(\alpha_s)$ corrections: for $2\rightarrow 5$, the 
BlackHat group has recently reported a result in Ref.~\cite{blkht}; for $2\rightarrow 4$, there are many published results, some of which involve
automation -- see Ref.~\cite{gslm} for a good review. What is the value of $\Delta\sigma_{\text{th}}$ for these impressive results?\par
\begin{figure}[h]
\begin{center}
\includegraphics[width=110mm]{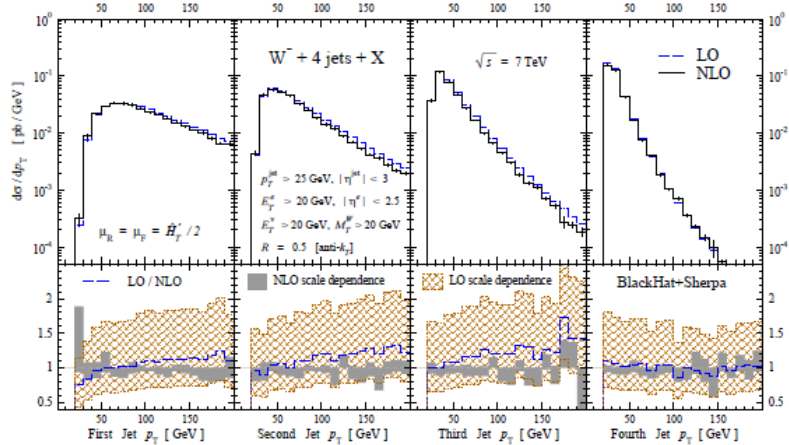}
\end{center}
\caption{\baselineskip=11pt  Results on the jet $p_T$ spectra in $W+4 jets+X$ production at the  LHC as reported in Ref.~\cite{blkht} and reviewed in part in Ref.~\cite{gslm}. The calculation is NLO in the leading color approximation for the virtual corrections.}
\label{fig-blkht}
\end{figure}

To illustrate the situation, we show in Fig.~\ref{fig-blkht} the application of
the result in Ref.~\cite{blkht} for the W+4jets+X process at LHC for the $p_T$
spectra of the leading 4 jets. The theoretical uncertainty of the NLO result is at the 10\% level, optimistically, if we only use the scale dependence variation. For precision
LHC applications, we need to know true value of $\Delta\sigma_{\text{th}}$, including the contributions of the PDF uncertainty, the technical precision uncertainty, ... .\par

\subsection{$\Delta\sigma_{\text{th}}$ in LHC Physics}

The basic paradigm in which we need to be able to prove
the value of $\Delta\sigma_{\text{th}}$ is the following one: we have, when arbitrary detector cuts
are allowed, the combination 
\begin{equation}
\text{MC}\cup \text{NLO}\cup \text{NNLO/NNLL},
\label{eq-prdgm}
\end{equation}
where the EW and mixed EW$\otimes$QCD corrections~\cite{qced,ncrsni,ew-qcd}
at the corresponding 
precision level are included here. In addition, this means that the 
quark masses $m_q\ne 0$
are general required for ISR at ${\cal O}(\alpha_s^n),\; n\ge 2$ so that an approach such as that in Ref.~\cite{ward-prd08} is needed.\par
\begin{figure}[h]
\begin{center}
\includegraphics[width=100mm]{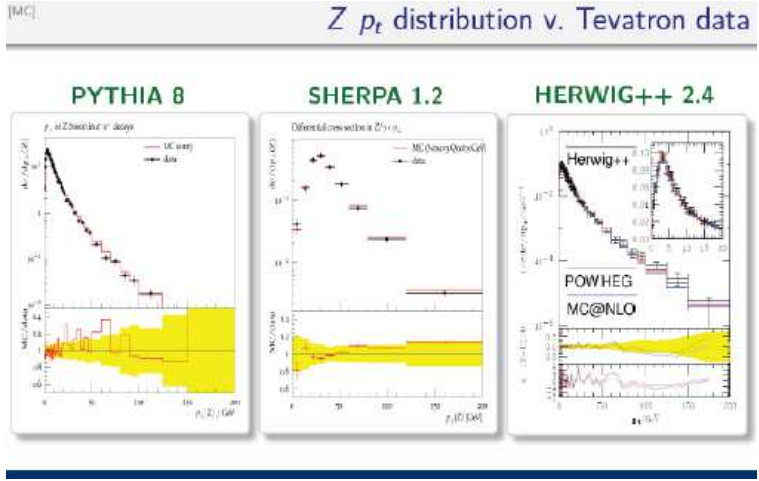}
\end{center}
\caption{\baselineskip=11pt  Results on $Z+X$ production at the 
Tevatron as reviewed in Ref.~\cite{gslm}.}
\label{fig-gslm3}
\end{figure}

There are by now some very standard tools available in the paradigm.The MC
is generically one of parton shower type~\cite{sjostrand-bk}, where in the
traditional FORTRAN
we have reference to Herwig6.5~\cite{hrwg} and Pythia6.4~\cite{pyth6}
and, more recently, in C++ we have Sherpa~\cite{shpa}, Herwig++~\cite{hrwg++} and Pythia8~\cite{pyth8}, to be specific. Again, we can get an estimate
of the values of $\Delta\sigma_{\text{th}}$ from Fig.~\ref{fig-gslm3}
where we see that the MC's Pythia8, Sherpa1.2 and Herwig++2.4 have 
uncertainties that very from 10\% to 50\% depending on the value of the
Z $p_T$ in single Z production at the Tevatron. This shows that the exact 
NLO corrections are necessary for precision studies 
already at the Tevatron.\par
Indeed, the parton shower MC's with exact ${\cal O}(\alpha_s)$ corrections, 
MC@NLO~\cite{mcnlo} and Powheg~\cite{pwhg},
provide realizations of the needed exact
NLO corrections. The resulting improvement in the
value of the $\Delta\sigma_{\text{th}}$ when including these exact corrections
can also be seen in Fig.~\ref{fig-gslm3} where the comparison of the variation
of the Z $p_T$ spectrum between MC@NLO and Powheg is given -- it reduces the
variation between the theory and the data considerably and shows a difference
between the two theory predictions at the level of $\sim 5-10\%$.
This is definite improvement but still leaves us quite a bit of work to reach
our goal in this process. We stress that while the two calculations in
MC@NLO and Powheg have both the exact ${\cal O}(\alpha_s)$ corrections, 
they differ in the corrections at ${\cal O}(\alpha_s^n), \; n\ge 2$.\par
To control all aspects of the contributions to $\Delta\sigma_{\text{th}}$
it is important for the observables to be infrared safe, as it is well-known.
The new infrared-safe anti-$k_T$ jet algorithm of Ref.~\cite{gslm2} allows
us to have infrared-safe jet definition in practice at the LHC, as it is apparently being adopted by the LHC collaborations~\cite{gslm}. The basic idea~\cite{gslm2}
is that one repeatedly combines pairs of objects with the smallest values of
$d_{ij}=R^2_{ij}/\text{max}(k^2_{Ti},k^2_{Tj})$ 
where $R_{ij}$ is an appropriately
normalized ``distance'' between the two objects i,j. This yields cones
in an infrared safe manner~\cite{gslm,gslm2}.\par
Further results on realizing exact ${\cal O}(\alpha_s)$ with n jets obtain:
the MENLOPS~\cite{menlops} project with NLO $Z$ production 
and LO $Z$ + $n$ jets + parton showers (using the CKKW and MLM merging
methods~\cite{merging}) adds in the multi-leg 
corrections. The combination
of NLO $Z$ and NLO $Z/\gamma$ + $n$ jets with parton showers has been done 
in Ref.~\cite{alioli-1}, for $n=1$. In all cases, one needs to prove one knows 
the corresponding values
of $\Delta\sigma_{\text{th}}$.\par
What is needed here is an NNLO with (resummed) parton shower MC for 
the complete realization of the LHC discovery potential. One needs 
specifically resummation
of all large collinear effects, so that one can use DGLAP-CS~\cite{dglap,cs} 
evolution in which $p_T$ is integrated out and one can use evolutions in 
which $p_T$ is alive~\cite{jadskrp}. One needs as well resummation of 
all large soft effects, including the Regge limit~\cite{eml,guido}, so that 
one needs these effects in both the collinear regime and in the non-collinear 
regime. One needs exact treatment of differential distributions through 
NNLO, with exact phase space and no miss-counting of efforts, including 
the effect of non-zero quark masses. The goal is event-by-event realization 
of these effects with exclusive exact NNLO with parton showers to yield 
a proof of the value of $\Delta\sigma_{\text{th}}$.\par
\begin{figure}[h]
\begin{center}
\includegraphics[width=100mm]{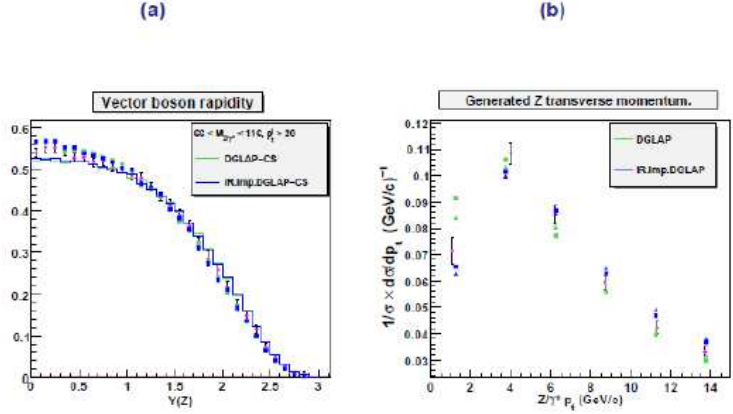}
\end{center}
\caption{\baselineskip=11pt  From Ref.~\cite{herwiri}, comparison with FNAL data: (a), CDF rapidity data on
($Z/\gamma^*$) production to $e^+e^-$ pairs, the circular dots are the data, the green(blue) lines are HERWIG6.510(HERWIRI1.031); 
(b), D0 $p_T$ spectrum data on ($Z/\gamma^*$) production to $e^+e^-$ pairs,
the circular dots are the data, the blue triangles are HERWIRI1.031, the green triangles are HERWIG6.510. In both (a) and (b) the blue squares are MC@NLO/HERWIRI1.031, and the green squares are MC@NLO/HERWIG6.510, where MC@NLO/X denotes the realization by MC@NLO of the exact ${\cal O}(\alpha_s)$ correction for the generator X. These are untuned theoretical results.}
\label{fig-hwri1}
\end{figure}

There is some progress in this effort as well. In (\ref{bscfrla}), resummation 
of collinear evolution is realized in the evolution of the $\{F_j\}$ and 
soft resummation(non-collinear) is realized in the calculation of 
$d\hat\sigma_{\text{res}}$. For example, from Ref.~\cite{qced} we have the 
representation
{\small
\begin{eqnarray}
&d\hat\sigma_{\rm res} = e^{\rm SUM_{IR}(QCED)}
   \sum_{{n,m}=0}^\infty\frac{1}{n!m!}\int\prod_{j_1=1}^n\frac{d^3k_{j_1}}{k_{j_1}} \cr
&\prod_{j_2=1}^m\frac{d^3{k'}_{j_2}}{{k'}_{j_2}}
\int\frac{d^4y}{(2\pi)^4}e^{iy\cdot(p_1+q_1-p_2-q_2-\sum k_{j_1}-\sum {k'}_{j_2})+
D_\rQCED} \cr
&\tilde{\bar\beta}_{n,m}(k_1,\ldots,k_n;k'_1,\ldots,k'_m)\frac{d^3p_2}{p_2^{\,0}}\frac{d^3q_2}{q_2^{\,0}},
\label{subp15b}
\end{eqnarray}}\noindent
where the new YFS-style~\cite{yfs} residuals
$\tilde{\bar\beta}_{n,m}(k_1,\ldots,k_n;k'_1,\ldots,k'_m)$ have $n$ hard gluons and $m$ hard photons and we show the final state with two hard final
partons with momenta $p_2,\; q_2$ specified for a generic $2f$ final state for
definiteness. The infrared functions ${\rm SUM_{IR}(QCED)},\; D_\rQCED\; $
are defined in Refs.~\cite{qced,irdglap1,irdglap2}. This  
simultaneous resummation of QED and QCD large IR effects is exact. Moreover,
the residuals $\tilde{\bar\beta}_{n,m}$ allow a rigorous parton 
shower/ME matching via their shower-subtracted 
counterparts $\hat{\tilde{\bar\beta}}_{n,m}$~\cite{qced}. When the 
formula in (\ref{subp15b}) is applied to the
calculation of the kernels, $P_{AB}$, in the DGLAP-CS theory itself, 
we get an improvement
of the IR limit of these kernels, an IR-improved DGLAP-CS theory~\cite{irdglap1,irdglap2} in which large IR effects are resummed for the kernels themselves.
The resulting new resummed kernels, $P^{exp}_{AB}$ as given in Ref.~\cite{irdglap1,irdglap2} and as illustrated below, yield a new resummed scheme for the PDF's and the reduced cross section: 
\begin{equation}
\begin{split}
F_j,\; \hat\sigma &\rightarrow F'_j,\; \hat\sigma'\; \text{for}\nonumber\\
P_{gq}(z)&\rightarrow P^{\text{exp}}_{gq}(z)=C_FF_{YFS}(\gamma_q)e^{\frac{1}{2}\delta_q}\frac{1+(1-z)^2}{z}z^{\gamma_q}, \text{etc.},
\end{split}
\end{equation}
with the same value for $\sigma$ in (\ref{bscfrla}) with improved MC stability
as discussed in Ref.~\cite{herwiri}. Here, the YFS~\cite{yfs} infrared factor 
is given by $F_{YFS}(a)=e^{-C_Ea}/\Gamma(1+a)$ where $C_E$ is Euler's constant
and we refer the reader to Ref.~\cite{irdglap1,irdglap2} for the definition of the infrared exponents $\gamma_q,\; \delta_q$. $C_F$ is the quadratic Casimir invariant for the quark color representation. The new MC Herwiri1.031~\cite{herwiri} gives the first realization of the new IR-improved kernels in the Herwig6.5~\cite{hrwg} environment. We illustrate it in comparison with Herwig6.510, both with and without
the MC@NLO exact ${\cal O}(\alpha_s)$ correction, in Fig.~\ref{fig-hwri1} in relation 
to D0 data~\cite{d0pt} on the Z boson $p_T$ in single Z production 
and the CDF data~\cite{galea} on the Z boson rapidity in the same process all at 
the Tevatron. We see~\cite{herwiri} 
that the IR improvement improves the $\chi^2/d.o.f$ 
in comparison with the data in both cases for the soft $p_T$ data and that
for the rapidity data it improves the $\chi^2/d.o.f$ before the application
of the MC@NLO exact ${\cal O}(\alpha_s)$ correction 
and that with the latter correction
the $\chi^2/d.o.f$'s are statistically indistinguishable. More importantly, 
this theoretical paradigm can be systematically improved in principle to reach any desired $\Delta\sigma_{\text{th}}$. The suggested accuracy at the 10\% level shows
the need for the NNLO extension of MC@NLO, in view of our goals
for this process. \par
\begin{figure}[h]
\begin{center}
\includegraphics[width=100mm]{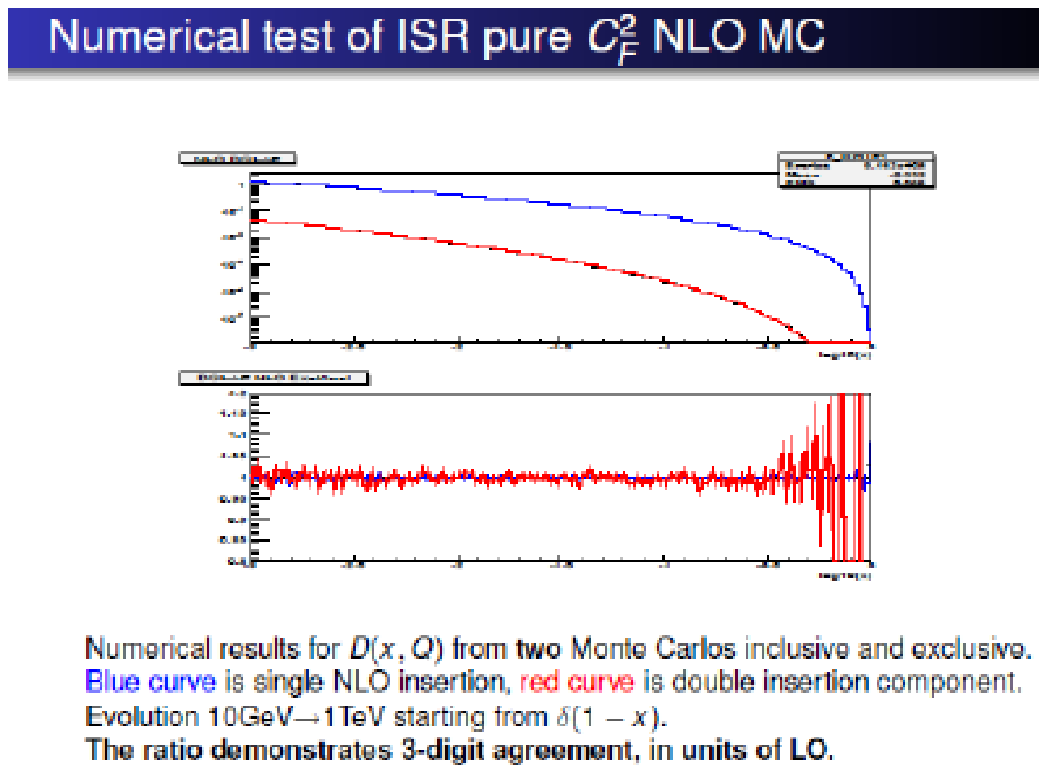}
\end{center}
\caption{\baselineskip=11pt  Numerical cross check of the approach in Ref.~\cite{jadskrp}.}
\label{fig-sk-jad1}
\end{figure}

We also note 
the developments in Refs.~\cite{jadskrp} aimed at exclusive realization
of the NLO correction to the DGLAP-CS kernels $P_{AB}$. We show in 
Fig.~\ref{fig-sk-jad1} numerical results that demonstrate the proof of concept 
for the non-singlet analysis as reported in Refs.~\cite{jadskrp} for the 
case that one NLO insertion
is added anywhere in the standard LL ladder representation of the solution 
for the respective distribution function. 
In this approach the modifications to the usual LL ladder for the respective
distribution function $\bar D_B$ can be seen in the formula{\small
\begin{equation}
\begin{split}
\bar D^{[1]}_B(x,Q) &= e^{-S_{ISR}}\Big\{\delta_{x=1}+ \sum_{n=1}^\infty \left(\prod_{i=1}^n\int_{Q>a_i>a_{i-1}}d^3\eta_i\rho^{(1)}_{1B}(k_i)\right){\Big[}\sum_{p=1}^n\beta^{(1)}_0(z_p)\\
&\;\; +\sum_{p=1}^n\sum_{j=1}^{p-1}W(\tilde{k}_p,\tilde{k}_j){\Big]}\delta_{x=\prod_{j=1}^nx_j}\Big\}
\end{split}
\label{sk-jad}
\end{equation}}
where the residuals $\beta^{[1]}_0$ and $W$ allow one to include the exclusive
effects for the NLO correction to the usual ladder solution, as expounded in Refs.~\cite{jadskrp}, where the Sudakov exponent $S_{ISR}$ and the real emission kinematics in (\ref{sk-jad}) are all defined. Similar results have been obtained for FSR. 
The next step is to add more
NLO insertions, 2,3, and so on. This is in progress. This theoretical paradigm
can in principle also be systematically improved to a given value of $\Delta\sigma_{\text{th}}$.\par
We then can prescribe a future QCD for the LHC as follows: it needs
exact amplitude-based resummation with NNLO hard corrections $({\cal O}(\alpha_s^2,\alpha\alpha_s,\alpha^2L^2))$ on an event-by-event basis via MC methods, with IR and collinearly improved showers, exact phase space and complete mass effects. The result will be provable control on $\Delta\sigma_{th}$ for LHC physics. In closing,
we thank Profs. S. Jadach and S. Yost and Dr. S. Majhi for useful discussions 
and we also thank Prof. Ignatios Antoniadis for the support and kind 
hospitality of the CERN TH Unit while this work was completed.\par

\end{document}